\documentclass[12pt,pre,aps,showpacs,preprint]{revtex4}

\usepackage{graphics}
\usepackage{graphicx}

\begin{document}

\title{Modeling Heterogeneous Materials via Two-Point Correlation Functions: II. Algorithmic Details and Applications}

\author{Y. Jiao}


\affiliation{Department of Mechanical and Aerospace Engineering,
Princeton University, Princeton New Jersey 08544, USA}

\author{F. H. Stillinger}


\affiliation{Department of Chemistry, Princeton University,
Princeton New Jersey 08544, USA}

\author{S. Torquato}

\email{torquato@electron.princeton.edu}

\affiliation{Department of Chemistry, Princeton University,
Princeton New Jersey 08544, USA}

\affiliation{Princeton Institute for the Science and Technology of
Materials, Princeton University, Princeton New Jersey 08544, USA}

\affiliation{Program in Applied and Computational Mathematics,
Princeton University, Princeton New Jersey 08544, USA}

\affiliation{Princeton Center for Theoretical Physics, Princeton
University, Princeton New Jersey 08544, USA}

\date{\today}

\begin{abstract}

In the first part of this series of two papers, we proposed a theoretical formalism
 that enables one to model and categorize heterogeneous materials (media) via
two-point correlation functions $S_2$ and introduced an efficient heterogeneous-medium
 (re)construction algorithm called the ``lattice-point'' algorithm. Here we
discuss the algorithmic details of the lattice-point procedure and
an algorithm modification using surface optimization to further
speed up the (re)construction process. The importance of the error
tolerance, which indicates to what accuracy the media are
(re)constructed, is also emphasized and discussed. We apply the
algorithm to generate three-dimensional digitized realizations of
a Fontainebleau sandstone and a boron carbide/aluminum composite
from the two-dimensional tomographic images of their slices through
the materials. To ascertain whether the information contained in
$S_2$ is sufficient to capture the salient structural features, we
compute the two-point cluster functions of the media, which are
superior signatures of the microstructure because they incorporate
topological connectedness information. We also study the reconstruction of
a binary laser-speckle pattern in two dimensions, in which the
algorithm fails to reproduce the pattern accurately. We conclude
that in general reconstructions using $S_2$ only work well for
heterogeneous materials with single-scale structures. However,
two-point information via $S_2$ is not sufficient to accurately
model multi-scale random media. Moreover, we construct realizations of
hypothetical materials with desired structural characteristics
obtained by manipulating their two-point correlation functions.

\end{abstract}


\maketitle

\section{Introduction}

Random heterogeneous multiphase materials or media are ubiquitous.
Examples include composites, porous media, biological
materials as well as cosmological structures, and their macroscopic properties are
of great interest \cite{1torquato}. In the first part of this series of two papers
\cite{yjiao} (henceforth referred to as paper I), we proposed a theoretical
formalism to model and categorize heterogeneous materials via two-point correlation
functions $S_2({\bf r})$, which can be interpreted as the probability of finding two
 points separated by the the dsiplacement vector ${\bf r}$ in one of the phases
\cite{1torquato}. In particular, we introduced the idea of the two-point correlation
 function space and its basis functions. In general, $S_2$ of a medium can be
expressed by a map $\wp$ on the associated basis functions, which is composed of
convex-combination and product operations. We also suggested a set of basis functions
 by examining certain known realizable analytical two-point correlation
functions. Moreover, we introduced an efficient isotropy-preserving  $S_2$-sampling algorithm,
namely the \textit{lattice-point} algorithm, but left the details of
the algorithm for another paper.

Here we will provide algorithmic details of the lattice-point methodology and
consider several nontrivial applications to illustrate the practical utility of our
theoretical formalism. In particular, we will apply the lattice-point
algorithm to generate three-dimensional (3D) digitized realizations of a
Fontainebleau sandstone and a boron carbide/aluminum composite from two-dimensional
(2D) tomographic images of their slices through the materials. To justify whether
the reconstructions are successful, one should also measure other statistical
descriptors of the media \cite{Yeong}. Here we compute the two-point cluster
function $C_2$ \cite{clusterII} (see definition and discussions in Sec.~IV),
which contains nontrivial ``connectedness'' information of the phases of interest.
By comparing $C_2$ of the target and reconstructed media, we demonstrate that
$S_2$ is indeed sufficient to capture the salient structural features in these
cases. We also study the reconstruction of a binary laser-speckle pattern in two
dimensions, and find that the algorithm fails to reproduce the target pattern accurately.
 We conclude that in general reconstructions using $S_2$ only work well for
heterogeneous materials with single-scale structures (those having only one
characteristic length scale). However, two-point information via $S_2$ is not
sufficient to accurately model multi-scale media (those composed of structural
elements associated with multiple characteristic length scales). Moreover, we will
show how the proper convex combinations of basis functions enable one to obtain
two-point correlation functions with desired properties and thus enable one to
generate a variety of structures with controllable morphological features. Note
that here we will mainly focus on modeling heterogeneous materials in three
dimensions, which complements the two-dimensional examples that we considered in
paper I.

Yeong and Torquato formulated the (re)construction problem as an 
energy-minimization problem using simulated annealing \cite{2PhysRevE.57.495}.
This has become a very popular (re)construction
technique \cite{ApplyA, ApplyB, ApplyC, ApplyD, Kumar}.
In this method, a nonnegative objective function $E$, called the ``energy,"
is defined as the sum of squared differences
between the target and sampled correlation functions. In our case, the energy
is given by

\begin{equation}
\label{eq201}
E = \sum_i[{S_2(r_i) - \hat{S}_2(r_i)}]^2,
\end{equation}

\noindent where $\hat{S}_2(r)$ and $S_2(r)$ is the target and sampled two-point
correlation function, respectively. An important issue in the 
(re)constructions based on the method of simulated
annealing is the choice of the \textit{energy threshold} $E_{th}$,
i.e., the error tolerance of discrepancies between the correlation functions of
generated medium and the imposed ones. When the
energy of the (re)constructed medium is below $E_{th}$, the (re)construction
process is terminated. The energy threshold is a key indicator of how accurately
the medium is (re)constructed. In our previous work, $E_{th}$ was always chosen to
be a very small number, e.g., $10^{-6}$; however, no quantitative analysis on why
such a value should be chosen was given. In this paper, we will discuss in detail
the significance of the energy threshold for both the orthogonal $S_2$-sampling
algorithm (only sampling $S_2$ along convenient orthogonal directions)
\cite{2PhysRevE.57.495, Yeong} and the lattice-point algorithm.

In the lattice-point procedure, we consider the
digitized medium (pixel system) to be a ``lattice-gas" system, in which the black pixels
behave like nonoverlapping ``particles'' moving from one lattice site to another.
The two-point correlation function of the medium is obtained by binning all the distances
 between the black pixels and dividing the number in each bin by the total number of
distances between all lattice sites within that bin. To generate a trial configuration,
a randomly selected ``particle'' (black pixel) is given a random displacement
subjected to the nonoverlapping constraint. Only the distances between the moved
``particle'' and all the others need to be recomputed in order to obtain $S_2$ of the trial
configuration. In this way, all directions in the digitized medium are effectively sampled;
moreover, the complexity of the algorithm is linear in $N_B$, i.e., the number of
nonoverlapping particles.

For heterogeneous materials containing well-defined inclusions, we will show that the computational
speed of the (re)construction process can be increased by an algorithm modification
using surface optimization, which is essentially a biased pixel-selection procedure.
 This algorithm modification is based on the fact that in the later stages of
(re)constructions, further refinements of the configurations are achieved by the
moves of black pixels on the surfaces of formed pixel-clusters. Thus, only the
``surface pixels'' are selected and given random moves to generate trial
configurations, i.e., the surfaces are optimized. The physical analog of this
process is solidification, in the later stage of which the re-arrangements of
solutes only occur on the surfaces of nucleated particles. We will see in the
following that one can obtain a much lower error (discrepancies between the
correlation functions of the (re)constructed medium and the imposed ones) when
surface optimization is properly applied. This improvement enables one to
quantitatively study the non-uniqueness problem of the (re)constructions
\cite{2PhysRevE.57.495, 4cule:3428, 5sheehan:53, Utz, Homometric, Vision}.

The rest of the paper is organized as follows: In Sec.~II, we describe the
lattice-point algorithm and the algorithm modification using surface optimization
in great detail. The choice of different pixel-lattices is also discussed. In
Sec.~III, we discuss the significance of the energy threshold for both the
orthogonal $S_2$-sampling method and the lattice-point algorithm. In Sec.~IV, we
suggest a general form of the map $\wp$ and apply the theoretical formalism to
model a Fontainebleau sandstone, a boron carbide/aluminum composite and a binary
laser-speckle pattern. We also show how to construct materials with
structural properties of interest by manipulating the parameters in the basis
functions and choosing proper combination coefficients. In Sec.~V, we make
concluding remarks.

\section{Algorithms for Generating Heterogeneous Materials}

In paper I, we derived the exact algebraic equations for the
(re)construction problem and showed that the equations have an
infinite number of solutions and cannot be solved rigorously. In
principle, the Yeong-Torquato scheme enables one to obtain one of
the solutions efficiently. The most time-consuming steps of the
scheme are the samplings of the two-point correlation function at
every trial configuration. An efficient $S_2$-sampling method
would dramatically speed up the (re)construction process.
Furthermore, an isotropy-preserving algorithm is required in
(re)constructions whenever a radial two-point correlation function
$S_2(r)$ ($r\equiv |{\bf r}|$) is employed for the case of
statistically homogeneous and isotropic media.

\subsection{Lattice-Point Algorithm}

\begin{figure}[bthp]
\begin{center}
$\begin{array}{c@{\hspace{2cm}}c}\\
\includegraphics[height=5cm,keepaspectratio]{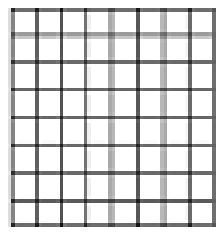} &
\includegraphics[height=5cm,keepaspectratio]{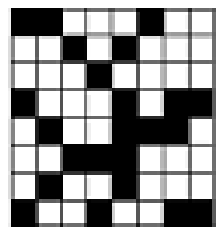} \\
\mbox{\bf (a)} & \mbox{\bf (b)}
\end{array}$
\end{center}
\caption{Digitized medium as lattice-gas system: (a) White pixels as unoccupied lattice sites. (b) Black pixels as non-overlapping ``gas molecules''.}
\label{fig1}
\end{figure}

The lattice-point algorithm is designed to sample the digitized representation of a
statistically homogeneous and isotropic medium in all possible directions
efficiently. For simplicity, we will illustrate the idea in 2D. Implementation of
the algorithm in three dimensions is a straightforward extension. Instead of
considering the digitized medium as a collection of black and white pixels, one can
think of the medium as a lattice-gas system: the black pixels are the ``gas
molecules'' and the white pixels are unoccupied lattice sites, as shown in
Fig.~\ref{fig1}. The ``gas molecules'' are free to move from one lattice site to
another, subject to the impenetrability condition, i.e., each lattice site can only
be occupied by one ``gas molecule''. Thus, the black pixels behave like hard
particles and the volume fraction of black phase is conserved during the evolution
of the system.

For a statistically homogeneous and isotropic particle system (e.g., an equilibrium
hard-sphere system or an equilibrium lattice-gas system in $d$-dimensional Euclidean
 space $\Re^d$), the most basic statistical descriptor of the spatial correlations
of the particles is the pair correlation function $g_2(r)$ \cite{1torquato}. The
quantity $\rho g_2(r)s_1(r)dr$ is proportional to the conditional probability of
finding the center of a particle in a $d$-dimensional spherical shell of volume
$s_1(r)dr$, given that there is another particle at the origin. Here $\rho = N/V$
is the number density of the system and $s_1(r)$ is the surface area of a $d$-dimensional
 sphere of radius $r$, which is given by

\begin{equation}
\label{eq1}
s_1(r) = \frac{2\pi^{d/2}r^{d-1}}{\Gamma(d/2)},
\end{equation}

\noindent where $\Gamma(x)$ is the Euler-Gamma function. Hence, for a finite system, 
integrating $\rho g_2(r)$ over the volume yields $(N-1)$, i.e., all the particles
except the one at the origin. Alternatively, $\rho g_2(r)s_1(r)dr$ is the average
number of particles at a radial distance between $r$ and $r+dr$ from a reference
particle. Practically, $g_2(r)$ can be obtained from simulations by generating a
histogram for the number of particles $n(r)$ contained in a concentric shell of
finite thickness (``bin'' width) $\Delta r$ at radial distance $r$ from a
arbitrarily chosen reference particle.

The two-point correlation function $S_2(r)$ of a statistically homogeneous and
isotropic medium can be interpreted as the probability that both of ends of a
randomly oriented line segment with length $r$ fall into the phase of interest,
say, the ``black'' phase. By comparison, we can see that the method of computing
$g_2(r)$ of an isotropic particle system implies a natural way of obtaining $S_2(r)$
 of a statistically isotropic digitized medium, which efficiently uses all possible
vector-information in the medium. For each black pixel (or ``gas molecule'') $i$,
the distances between pixel $i$ and all the other pixels $j$ are computed and binned
 to generate a histogram for the number of black pixels separated from each other
by distance $r$. Another histogram for the number of lattice sites separated from a
reference site by distance $r$ is also generated by computing and binning all
possible site-separation distances. Suppose the histograms for black pixels and
lattice sites are stored in the array $BN[r]$ and $SN[r]$, respectively. It is easy
to show that the two-point correlation function can be obtained from

\begin{equation}
\label{eq2}
S_2[r] = BN[r]/SN[r],
\end{equation}

\begin{figure}[bthp]
\begin{center}
$\begin{array}{c@{\hspace{2cm}}c}\\
\includegraphics[height=5cm,keepaspectratio]{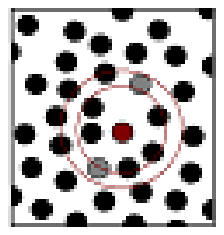} &
\includegraphics[height=5cm,keepaspectratio]{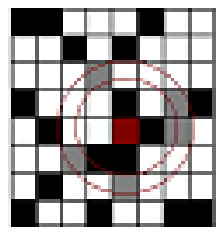} \\
\mbox{\bf (a)} & \mbox{\bf (b)}
\end{array}$
\end{center}
\caption{(color online). Centers of particles contained in the concentric shell of a
reference particle: (a) In a equilibrium hard-disk system. (b) In a lattice-gas
system (or a digitized medium).}
\label{fig2}
\end{figure}

\noindent (see Fig.~\ref{fig2}). In other words, to compute $S_2$ we first calculate
 the fraction of occupied lattice sites separated by distance $r$ ($r^2$ is an
integer) from a reference black pixel. Then we average this fraction over all black
pixels (by choosing every black pixel as reference pixel once) to obtain $S_2(r)$.
This procedure is consistent with the geometrical probability interpretation of
$S_2$. For the digitized media, a natural bin width could be the characteristic size
 of the pixel of the lattice, e.g., the edge length of a square pixel if square
lattice is used.

At each step in the simulated annealing process, a trial configuration is generated
by moving a randomly selected black pixel to an unoccupied lattice site.
A \textit{configuration array} can be used to speed up this process. In our 2D
implementation, the configuration array is a 2D array with the entries being 1 and
0, corresponding to the occupied and unoccupied lattice sites, respectively. When a
trial move is made, first we need to test whether it violates the impenetrability
condition by checking whether the underlying lattice site is occupied or not. This
process just requires constant access time to an entry of the configuration array.
Note that several trial moves can be attempted before a trial configuration is
found for a high-density system. After a trial configuration is generated, the old
position information of the selected pixel is stored in the array designated $PO$.

The next step is to recompute the two-point correlation function for the trial
configuration. To do this efficiently, a \textit{distance matrix} is set up when
the system is initialized and updated if a trial configuration is accepted. Note
that this matrix is symmetric. For each trial configuration, only the position of a
randomly selected black pixel is changed, e.g., the $k^{th}$ pixel. Thus, only the
$k^{th}$ row and column of the distance matrix need to be updated, which requires
$N_B$ operations ($N_B$ is the total number of black pixels in the system). The old
entries in the $k^{th}$ row (or column) of the distance matrix are stored in the
array $DO$. It is also unnecessary to recompute the entire array $BN[r]$. Recall
that $BN[r]$ contains the binned number of black pixels separated from each other
by distances between $r$ and $r+\Delta r$ ($\Delta r$ is the bin width). Once the
$k^{th}$ pixel is selected, the distances between pixel $k$ and all the other pixels
 are binned and stored in the array $BNO[r]$. After the trial configuration is
generated, the new distances between pixel $k$ and all the other pixels are binned
and stored in the array $BNN[r]$. The array $BN[r]$ is updated as follows for each
$r$:

\begin{equation}
\label{eq3}
BN[r] - BNO[r] + BNN[r] \rightarrow BN[r].
\end{equation}

\noindent $S_2[r]$ is then recomputed using Eq.~(\ref{eq2}) and the trial
configuration is accepted with the probability given by Eq.~(57) in paper I.
If the trial configuration is rejected, all information of the old configuration
can be restored easily using the arrays $PO$, $DO$, $BNO[r]$ and $BNN[r]$. For
example, $BN[r]$ of the old configuration can be restored by

\begin{equation}
\label{eq31}
BN[r] - BNN[r] + BNO[r] \rightarrow BN[r].
\end{equation}

The above procedures are repeated until the energy of the (re)constructed medium is
below the energy threshold [see Eq.~(\ref{eq201}) in Sec.~III] or the total number
of evolutions reaches the prescribed limit value.

\subsection{Algorithm Modification Using Surface Optimization}

In the (re)construction of media composed of well-defined ``particles'' or large
clusters, we find out that in the later stages of simulated annealing, the random
pixel-selection process is very inefficient. Many trial moves are rejected because
a majority of selected pixels are inside the formed ``particles'' or clusters and
it is energetically unfavorable to move them outside. This requires the use of a
biased pixel-selection process, i.e., surface optimization.

The idea of surface optimization is analogous to the physical process of
solidification: when the nuclei of proper sizes have been formed, they capture more
solutes from surrounding solution to further decrease the total free energy. In the
simulated annealing process, once the ``nuclei'' are formed, the random pixel-selection
process is replaced by a biased pixel-selection process, i.e., only those
in the surrounding ``solution'' or on the surface of a ``nucleus'' are selected to
be moved to or along the surface of randomly selected ``nuclei''.

It is natural and easy to incorporate the surface optimization with the lattice-point
algorithm because the black pixels are already considered as ``molecules''.
Each black pixel is assigned a ``free energy'', which is the minus of the number of
the nearest neighbors of that pixel. If a pixel is inside a ``nuclei'', it has the
largest number of nearest neighbors and the lowest ``free energy'', which equals
$-4$ if a square lattice is used and $-6$ if a triangular lattice is used. If the pixel
is on the surface of a ``nucleus'' or in the surrounding ``solution'', its ``free
energy'' will be relatively higher. The highest free energy is $0$, which means the
pixel is separated from all the others. Thus, the black pixels are grouped into two
subsets: the \textit{low-energy} subset that contains pixels with lowest ``free
energy'' and the \textit{high-energy} subset that contains the other pixels. When the
trial move is made, only the pixels in the high-energy subset are selected to bias
the move. If the trial move is accepted, the free energy of the moved pixel and its
neighbor pixels are recomputed; the subsets are updated.

\begin{figure}[bthp]
\begin{center}
$\begin{array}{c}\\
\includegraphics[height=6cm,keepaspectratio]{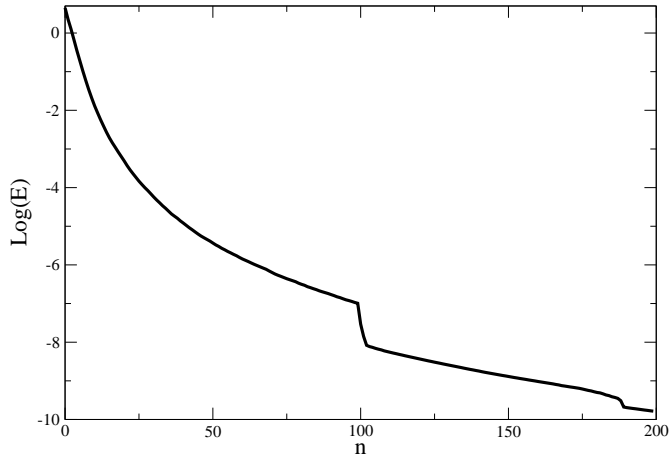}
\end{array}$
\end{center}
\caption{Energy $E$ of the constructed medium as a function of stages $n$. The
target autocovariance function is the Debye random medium function $f_D$ given by
Eq.~(\ref{eq101}), with $a=30$ (pixels) and volume fractions $\phi_1 = \phi_2 = 0.5$.
 The linear size of the system $N = 200$ (pixels). Surface optimization is applied
at $n=100$, when large clusters have been formed in the constructed
medium.}
\label{fig3}
\end{figure}

Numerical experiments show that applying surface optimization properly will further
decrease the final energy of the (re)constructed medium by a factor of $10^{-2}$
(see Fig.~\ref{fig3}), if the same cooling schedule is used. Thus, surface
optimization enables one to efficiently produce more accurate structures associated
with the imposed correlation functions. Moreover, by setting a lower energy
threshold, one can address the non-uniqueness issue quantitatively. Progress on this
topic will be reported in our future publications.

\subsection{Choice of Lattices}

\begin{figure}[bthp]
\begin{center}
$\begin{array}{c@{\hspace{2cm}}c}\\
\includegraphics[height=5cm,keepaspectratio]{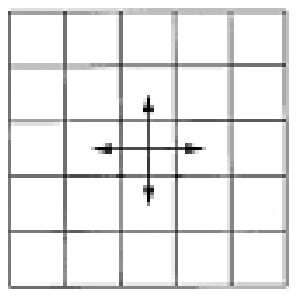} &
\includegraphics[height=5cm,keepaspectratio]{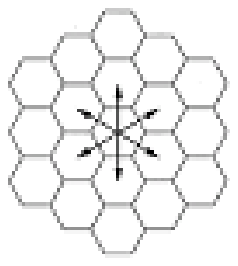} \\
\mbox{\bf (a)} & \mbox{\bf (b)}
\end{array}$
\end{center}
\caption{(a) The pixels (squares) and intrinsic directions of a square lattice. (b)
The pixels (hexagons) and intrinsic directions of a triangular lattice.}
\label{fig4}
\end{figure}

A 2D digitized heterogeneous material (medium) can be represented as a 2D array and
the real morphology of the material also depends on the lattice on which the system
of pixels is built \cite{yjiao}. Here, we mainly focus on two commonly used lattices
in the literatures, namely, the square lattice and the triangular lattice (see
Fig.~\ref{fig4}).

\begin{figure}[bthp]
\begin{center}
$\begin{array}{c@{\hspace{2cm}}c}\\
\includegraphics[height=5cm,keepaspectratio]{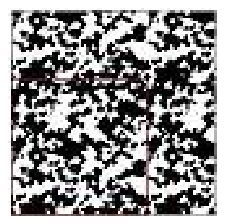} &
\includegraphics[height=5cm,keepaspectratio]{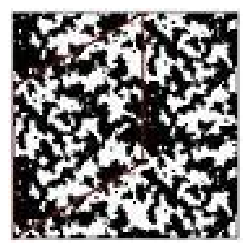} \\
\mbox{\bf (a)} & \mbox{\bf (b)}
\end{array}$
\end{center}
\caption{Constructions of media with large-scale structures. (a) Medium generated on a square
lattice with a square unit cell. (b) Medium generated on a triangular lattice with a rhombus unit cell.}
\label{fig5}
\end{figure}

Implementation of the algorithms discussed above on a square lattice is easier than
that on a triangular lattice because the former has orthogonal lattice vectors.
However, the pixels of a square lattice (squares) only have $4$-fold symmetry while
those of a triangular lattice (hexagons) have $6$-fold symmetry, as shown in
Fig.~\ref{fig4}. We find that for the media with long-range correlations or large-scale
structures, using a square lattice usually introduces undesirable anisotropy
in the (re)constructed systems, as shown in Fig.~\ref{fig5}; while for the media
without long-range order or large-scale structures, both lattices work well.

It is worth pointing out that the triangular lattice is superior to the square
lattice in the (re)construction of \textit{anisotropic} materials. In that case, one
cannot count on lattice-point algorithm, which only uses radial averaged structural
information. Several optimization directions need to be specified and processed
separately. The triangular lattice has six intrinsic directions due to the higher
symmetry of its pixel shape while the square lattice only has four (see Fig.~\ref{fig4}).
Methods using more optimization directions have been proposed by Torquato and
coworkers \cite{5sheehan:53}, i.e., along the $45$- and $135$-degree directions in
the square lattice; but one needs to pay the cost of complexity of implementation in
these cases.

\section{Significance of The Energy Threshold}

An important issue that has not been emphasized in our previous work on the
(re)construction algorithms is the choice of the \textit{energy threshold} $E_{th}$,
i.e., the error tolerance of discrepancies between the statistical properties of the
generated structure and the imposed ones. Recall that in our case, the \textit{energy} is defined
 as the sum of squared differences between target $\hat{S}_2(r)$ and $S_2(r)$ of the
constructed medium [see Eq.~(\ref{eq201})] and the energy threshold
$E_{th}$ is a prescribed value of $E$ such that
when $E \le E_{th}$, the (re)construction is terminated. $E_{th}$ indicates how
accurately the medium is (re)constructed. A smaller $E_{th}$ means the two-point
correlation function of the generated medium matches the imposed one better.

In the following, we will estimate the change of the energy of a digitized medium
caused by a perturbation of its original structure (i.e., displacing a randomly
selected black pixel). This change of energy can be considered as the energy
difference between the target and the constructed medium. Thus, by imposing a
particular value of the energy difference (i.e., $E_{th}$), we can in turn estimate
the number of perturbed black pixels. In this way, we quantitatively relate the
energy threshold to the question of how well the medium is (re)constructed. In
general, the energy difference depends on how $S_2$ is sampled, the linear size of
the system $N$ and the number of black pixels $N_B$ (volume fraction of black
phase $\phi_1$), which will be discussed accordingly.

\subsection{$E_{th}$ for the Lattice-Point Algorithm}

 First, we consider the energy threshold of the lattice-point algorithm. The
two-point correlation function of a digitized medium is computed from Eq.~(\ref{eq2}).
If the generated structure perfectly matches the target structure, the
energy given by Eq.~(\ref{eq201}) should be exactly 0. Suppose the perfect
(re)constructed structure is perturbed by moving one of its black pixels to an
unoccupied lattice site; then $S_2(r)$ is different from $\hat{S}_2(r)$ and $E$
becomes a small positive number. Note that we assume the perturbed structure does not
have the same $S_2$ as the original structure, i.e., there is no structural
degeneracy. Let $E_{min}$ denote the smallest positive value of $E$. From
Eq.~(\ref{eq201}), we see that the difference between $S_2$ and $\hat{S}_2$ of every bin
contributes to $E$. From Eq.~(\ref{eq2}) and the fact that the last bin (farthest
away from the reference center) contains the largest number of pairs of lattice
sites, we have

\begin{equation}
\label{eq801}
E_{min} = \left [{S_2\left(\left[\frac{N}{2}\right]\right) - \hat{S}_2\left(\left[\frac{N}{2}\right]\right)}\right ]^2 = \frac{4}{\omega_N^2},
\end{equation}

\noindent where $[N/2]$ is the integer part of $N/2$ and $\omega_N$ is the number of
elements of the integer set $\Omega_N$, which is defined by

\begin{equation}
\label{eq802}
\Omega_N = \left\{(m, n)~|~-\left [\frac{N}{2}\right] \le m, n \le \left [\frac{N}{2}\right], \left (\left[\frac{N}{2}\right]-1\right)^2 \le m^2+n^2 \le \left(\left [\frac{N}{2}\right]\right)^2  \right\}.
\end{equation}

\noindent In other words, Eq.~(\ref{eq801}) estimates $E_{min}$ using the change of
$E$ from 0 caused by removing (adding) one pair of black pixels from (to) the last
bin (the pixels bounded in pairs due to the symmetry of the distance matrix). Also note
that the pair of pixels are originally in (or moved to) a ``bin'' for the pair distance
larger than $[N/2]$, which we do not take into account for computation of $S_2$.

For a particular value of $E_{th}$, the maximum number of \textit{misplaced pairs} of
 black pixels can be estimated by

\begin{equation}
\label{eq803}
N_m = \frac{E_{th}}{E_{min}} = \frac{1}{4}\omega_N^2 E_{th}.
\end{equation}

\noindent The ratio of the number of misplaced pairs of black pixels over the total
number of pairs of black pixels is given by

\begin{equation}
\label{eq804}
\gamma = \frac{N_m}{N_{tot}} = \frac{\omega_N^2 E_{th}}{4 \phi_1^2 N^4}.
\end{equation}

\noindent Instead of specifying $E_{th}$, one can also specify the ratio
$\gamma = \gamma_s$; and the threshold can be computed from Eq.~(\ref{eq804}):

\begin{equation}
\label{eq806}
E_{th} = \frac{4}{\omega_N^2}\gamma_s\phi_1^2 N^4.
\end{equation}

For example, consider a system composed of $200 \times 200$ pixels ($N = 200$) and
$\phi_1 = 0.5$. The total number of lattice sites is $N_S = 40,000$ and the total
number of black pixels is $N_B = 20,000$. $\omega_N$ can be obtained numerically,
which is $\omega_N \sim 10^6$. From Eq.~(\ref{eq801}), we have
$E_{min} \sim 10^{-12}$. Suppose we choose the threshold $E_{th} = 10^{-9}$, from
Eq.~(\ref{eq803}), we have the number of misplaced pairs of black pixels
$N_m \sim 10^3$, which seems to be a large number. However, when considering the
total number of pairs in the system, we have $N_{tot} = N_B^2 = 4\times 10^{8}$;
from Eq.~(\ref{eq804}), the ratio is given by

\begin{equation}
\label{eq805}
\gamma = \frac{N_m}{N_{tot}} \simeq \frac{10^3}{4\times10^8} = 2.5\times10^{-6},
\end{equation}

\noindent which means only one out of a half million pairs is put in the wrong bin.
Thus, the medium is (re)constructed to a very high accuracy. In our simulations, we
choose the threshold $E_{th} = 10^{-9}$.

\subsection{$E_{th}$ for the Orthogonal $S_2$-Sampling Algorithm}

 For the orthogonal $S_2$-sampling algorithm, $S_2(r)$ is sampled line (column) by
line (column) by moving a line (column) segment of length $r$ one pixel distance
each time and counting the times that both ends of the segment are black pixels.
This number is then divided by the total number of times that one moves the line
(column) segment (total number of pixels on the line (column)) to obtain $S_2$ of
that line (column). Finally, the $S_2$ sampled from different lines and columns of
the digitized medium are averaged to compute the $S_2$ of the whole medium.
Similarly, consider the perfect (re)constructed structure is perturbed by moving
one of its black pixels to an unoccupied lattice site and there is no structural
degeneracy, $E_{min}$ is given by

\begin{equation}
\label{eq901}
E_{min} = \frac{1}{N^4},
\end{equation}

\noindent where $N$ is the linear size of the system. For a particular $E_{th}$,
the maximum number of \textit{misplaced black pixels} is given by

\begin{equation}
\label{eq902}
N_m = \frac{E_{th}}{E_{min}} = N^4 E_{th}.
\end{equation}

\noindent Thus, the ratio of misplaced black pixels over the total number of black
pixels can be obtained by

\begin{equation}
\label{eq903}
\gamma = \frac{N_m}{N_P} = \frac{1}{\phi_1}N^2 E_{th}.
\end{equation}

\noindent If stead, $\gamma = \gamma_s$ is specified, the required energy threshold
$E_{th}$ can be obtained from Eq.~(\ref{eq903}):

\begin{equation}
\label{eq904}
E_{th} = \frac{\gamma_s \phi_1}{N^2}.
\end{equation}

Consider the same system used in the previous section ($N = 200$, $\phi_1 = 0.5$).
If we require only 10 black pixels are misplaced, thus, the ratio $\gamma_s = 10/N^2
= 2.5\times 10^{-4}$. From Eq.~(\ref{eq904}), we have

\begin{equation}
\label{eq905}
E_{th} = \frac{2.5\times 10^{-4}\times 0.5}{4\times 10^4} \simeq 10^{-9}.
\end{equation}

\noindent In other words, if we choose the threshold $E_{th} = 10^{-9}$, only 10
black pixels in the system are misplaced. The medium is (re)constructed to a high
accuracy.

It is worth noting that in the above discussions, we assume that the perturbed
structure does not have the same $S_2$ of the original structure, i.e., there is no
structural degeneracy. In general, however, structural degeneracy does exist (i.e.,
media with the same $S_2$ but different $S_3$, $S_4$, ...). Results concerning these
non-uniqueness issues will be reported in our future publications.

\section{Applications of the Theoretical Formalism}

In this section, we illustrate the practical utility of our theoretical formalism in
both two and three dimensions. In particular, we will consider two kinds of
applications. Firstly, given a set of basis functions (may not be complete), one can
express the scaled autocovariance functions (two-point correlation functions) of a
statistically homogeneous and isotropic medium in terms of the specified basis
functions with certain accuracy. Realizations of the materials can be generated
using proper (re)construction procedures and subsequent analysis can be performed on
the images to obtain effective macroscopic properties of interest; see, e.g.,
Ref.~\cite{Kumar}. Secondly, the map $\wp$ (see the discussion below) enables one to
construct candidates of realizable two-point correlation functions with properties of
 interest, which in turn enables one to design and investigate materials with desired
 structural characteristics.

The (re)construction of realizations of 3D medium from the information obtained from
a 2D micrograph or image is of great value in practice \cite{Yeong}. Therefore, we
will apply the lattice-point algorithm to generate three-dimensional digitized
realizations of a Fontainebleau sandstone and a boron carbide/aluminum composite from
 two-dimensional tomographic images of their slices through the materials. In a
successful reconstruction, other deemed crucial structural characteristics besides
$S_2$ obtained from the reconstructed medium should also agree closely with that
of the target medium. Consequently, in order to judge quantitatively how well the
reconstructions are, we will measure and compare another important morphological
descriptor, i.e., the two-point cluster function $C_2({\bf x}_1, {\bf x}_2)$, defined
 to be the probability of finding two points at ${\bf x}_1$ and ${\bf x}_2$
 in the same cluster of the phase of interest \cite{clusterII}. For statistically
homogeneous and isotropic media, $C_2$ only depends on the
relative scaler distances between the points, i.e., $C_2({\bf
x}_1, {\bf x}_2) = C_2(|{\bf x}_1-{\bf x}_2|) = C_2(r)$. Note that
$C_2$ contains nontrivial topological ``connectedness''
information. When large clusters are present in the medium, $C_2$
becomes a long-ranged function and its integral will diverge if
the phase of interest percolates. The measurement of $C_2$ for a
3D material sample cannot be made from a 2D cross-section of the
material, since it is an intrinsically 3D microstructural function
\cite{1torquato}. To sample $C_2$ from a digitized medium, we
associate each
 pixel with a cluster-index that indicates to which cluster the pixel belongs, and
only bin the distances of pixel-pairs in the same cluster. Then the number of pair
distances in each bin is normalized by the total number of distances between the
lattice sites within that bin, which is similar to the procedure of sampling $S_2$
discussed in Sec.~II.

We will also study the reconstruction of a binary laser-speckle pattern in two
dimensions and show that the algorithm cannot reproduce the pattern accurately.
Moreover, we introduce and discuss a classification of heterogeneous materials into
\textit{multi-scale media} and \textit{single-scale media}.
A multi-scale medium (MSM) is the one in which there are multiple
characteristic length scales associated with different structural
elements. Examples of MSM include fractal patterns and
hierarchical laminate composites \cite{laminate}.
A single-scale medium (SSM) is the one composed of
structural elements associated with only one characteristic length
scale, such as a Fontainebleau sandstone and a boron
carbide/aluminum composite. We conclude that in
general reconstructions using $S_2$ only work well for heterogeneous materials with
single-scale structures. However, two-point information via $S_2$ is not sufficient
to capture the key structural features of multi-scale media. Before presenting the
aforementioned applications, we will first consider a general form of $\wp$, which
includes all possible convex combinations of the basis functions.

\subsection{A General Form of $\wp$}

In paper I, we introduced the idea of expressing the two-point correlation function
of a statistically homogeneous and isotropic medium through a selected set of bases
of the two-point correlation function space. In practice, it is convenient to use the
 scaled autocovariance functions that are equivalent to the two-point correlation
functions, which are defined as follows:

\begin{equation}
\label{eq203}
f(r) \equiv \frac{S^{(i)}_2(r) - \phi^2_i}{\phi_1\phi_2}.
\end{equation}

\noindent Suppose $\{f_i(r)\}_{i=1}^m$ is a set of bases of scaled autocovariance
functions and $\wp$ is an map on $\{f_i(r)\}_{i=1}^m$ composed of convex combinations
 and products of $f_i(r)$ ($i=1,...,m$), thus

\begin{equation}
\label{eq4}
f(r) = \wp[\{f_i(r)\}_{i=1}^m] = \wp[f_1(r),f_2(r),...,f_m(r)],
\end{equation}

\noindent is also a realizable scaled autocovariance function. Note that the choice
of basis functions is not unique.

In general, one can consider that $\wp[\{f_i(r)\}_{i=1}^m]$ takes the form

\begin{equation}
\label{eq7}
\wp[\{f_i(r)\}_{i=1}^m] = \sum\limits_i \alpha_i f_i(r) + \sum\limits_{i,j} \beta_{ij} f_i(r)f_j(r) + \cdots,
\end{equation}

\noindent where the coefficients satisfy the condition

\begin{equation}
\label{eq8}
\sum\limits_i \alpha_i + \sum\limits_{i,j} \beta_{ij} + \cdots = 1.
\end{equation}

\noindent Then one can use standard regression methods to obtain the set of
coefficients such that $f(r) = \wp[\{f_i(r)\}_{i=1}^m]$ is the best approximation of
the target function pointwisely.

The basis functions we consider here include Debye random medium
function $f_D(r)$, the family of polynomial functions
$f_P^{(n)}(r)$, damped oscillating function $f_O(r)$,
overlapping-sphere function $f_S(r)$ and symmetric-cell material
function $f_C(r)$, which are discussed in paper I.

\subsection{Modeling Real Materials}

\subsubsection{Fontainebleau Sandstone}

\begin{figure}[bthp]
\begin{center}
$\begin{array}{c}\\
\includegraphics[width=5cm,keepaspectratio]{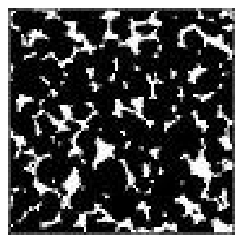}\\
\mbox{\bf (a)} \\\\\\
\includegraphics[height=5.5cm,keepaspectratio]{fig6b.eps}\\
\mbox{\bf (b)}
\end{array}$
\end{center}
\caption{(a) A microstructrual image of a slice of a Fontainbleau sandstone
\cite{sandstone3}. The black phase is the solid phase with $\phi_1 = 0.825$, and the
white phase is the void phase with $\phi_2 = 0.175$. (b) The two-point correlation
function $S^{FS}_2(r)$ of the void (white) phase.}
\label{fig6}
\end{figure}

First, we investigate structural properties of a Fontainebleau sandstone from a
two-dimensional tomographic image of a slice through the material sample. Sandstone
is an important porous medium in geo-physical and petroleum applications and has been
 the focus of many studies \cite{sandstone3, sandstone2, sandstone, Yeong}. A
microstructural image of a slice of a Fontainebleau sandstone is shown in
Fig.~\ref{fig6}(a), in which the black areas are solid phases (phase 1) and the white
 areas are void phases (phase 2). The two-point correlation function of the
\textit{void} phase $S_2^{FS}(r)$ is shown in Fig.~\ref{fig6}(b).

\begin{figure}[bthp]
\begin{center}
$\begin{array}{c@{\hspace{2cm}}c}\\
\includegraphics[height=6cm,keepaspectratio]{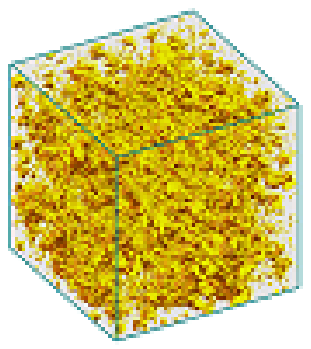} &
\includegraphics[height=5cm,keepaspectratio]{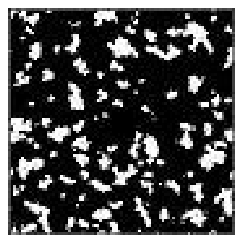} \\
\mbox{\bf (a)} & \mbox{\bf (b)}
\end{array}$
\end{center}
\caption{(color online). (a) The reconstructed 3D realization of the Fontainbleau
sandstone from $S^{FS}_2(r)$ of the void phase. The void phase is shown in yellow and
 the solid phase is transparent for easy visualization. (b) A 2D slice of the
constructed 3D realization. The solid phase is shown in black and the void phase is
shown in white. The linear size of the system $N = 160$ (pixels).}
\label{fig7}
\end{figure}

The scaled autocovariance function of the void phase in Fontainebleau sandstone
$f^{FS}(r)$ can be approximated by the convex combination of $f_D(r)$ and $f_O(r)$ as
 follows:

\begin{equation}
f^{FS}(r) = \alpha_1 f_D(r) + \alpha_2 f_O(r),
\label{eq5}
\end{equation}

\noindent where $\alpha_1 = 0.77$, $\alpha_2 = 0.23$ are the combination coefficients
 and

\begin{equation}
f_D(r) = \exp(-r/a),
\label{eq101}
\end{equation}

\begin{equation}
f_O(r) = \exp(-r/b)\cos(qr+\psi),
\label{eq102}
\end{equation}

\noindent where $a = 3~(pixel)$, $b = 6.5~(pixel)$ are the effective correlation
length; $q = 0.2~(pixel^{-1})$ is the oscillating frequency and $\psi = 0$ is the
phase angle. The two-point correlation function $S_2^{FS}(r)$ is approximated by

\begin{equation}
\label{eq6}
S_2^{FS}(r) = f^{FS}(r)\phi_1\phi_2 + \phi_2^2 + \delta S_2(r),
\end{equation}

\noindent where $\phi_1 = 0.825$, $\phi_2 = 0.175$ are volume
fractions of the solid and void phase, respectively; $\delta
S_2(r)$ is the discrepancy between the sampled two-point
correlation function and the basis-function approximation. The
average of the absolute values of discrepancies $\Delta S_2$,
which indicates how well the sampled two-point correlation
function is approximated by the convex combination, is defined as:

\begin{equation}
\label{eq601}
\Delta S_2 = \frac{1}{N_L}\sum_r \left |\delta S_2(r)\right| \simeq 5.2\times10^{-4}.
\end{equation}

\noindent where $N_L = 80$ is the sample-length. Note that
although the general form of $\wp$ given by Eq.~(\ref{eq7}) would
work well if a complete set of basis functions is given, in the
present case, our practical approach is enough.

\begin{figure}[bthp]
\begin{center}
$\begin{array}{c}\\
\includegraphics[height=5.5cm,keepaspectratio]{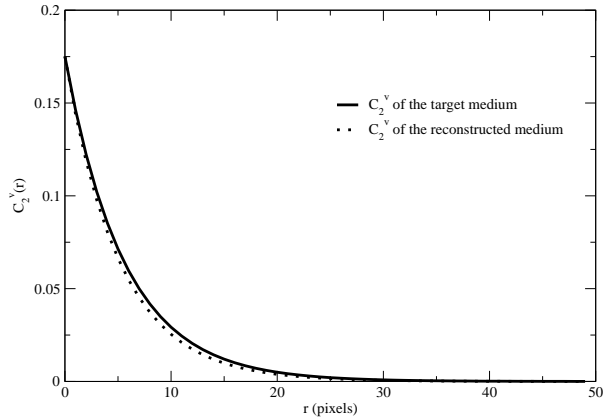}\\
\end{array}$
\end{center}
\caption{The two-point cluster functions $C^v_2(r)$ for the void (white) phase of the
 target and reconstructed 2D slices of the Fontainebleau sandstone.}
\label{fig71}
\end{figure}

The 3D reconstruction of the Fontainebleau sandstone from $S_2^{FS}(r)$ obtained from
 the digitized image of a 2D slice [Fig.~\ref{fig6}(a)] is shown in Fig.~\ref{fig7}.
Visually, the reconstruction provides a good rendition of the true sandstone
microstructure, as can be seen by comparing the 2D images. As pointed out earlier,
to ascertain whether the reconstruction is quantitatively successful, we also measure
 $C_2$ of the target and generated media. Since it is an intrinsically 3D
microstructural function, we only compute and compare the two-point cluster functions
 of the 2D slices. The measured $C^v_2(r)$ for the void phase of the target and the
reconstructed slices of the Fontainebleau sandstone are shown in Fig.~\ref{fig71}.
The figure reveals that although $C^v_2(r)$ of the generated medium is slightly below
 that of the target medium, the discrepancies are acceptable (e.g., the largest
discrepancy $|\delta C^v_2| \sim 5\times 10^{-3}$). Thus, we consider the
reconstruction is successful. Also note that the close agreement of the two-point
cluster function indicates that $C^v_2(r)$ is largely determined by $S^{FS}_2(r)$ of
the medium and suggests that $C^v_2(r)$ might be expressed as a functional of
$S^{FS}_2(r)$.

\subsubsection{Boron Carbide/Aluminum Composite}

\begin{figure}[bthp]
\begin{center}
$\begin{array}{c}\\
\includegraphics[width=5cm,keepaspectratio]{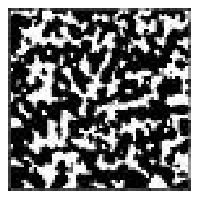}\\
\mbox{\bf (a)} \\\\\\
\includegraphics[height=5.5cm,keepaspectratio]{fig9b.eps}\\
\mbox{\bf (b)}
\end{array}$
\end{center}
\caption{(a) A digitized image of a boron carbide/aluminum composite \cite{ceramic}.
The black phase is boron carbide with $\phi_1 = 0.647$, and the white phase is
aluminum with $\phi_2 = 0.353$. (b) The two-point correlation $S^{Al}_2(r)$ of the
aluminum phase.}
\label{fig8}
\end{figure}

\begin{figure}[bthp]
\begin{center}
$\begin{array}{c@{\hspace{2cm}}c}\\
\includegraphics[height=6cm,keepaspectratio]{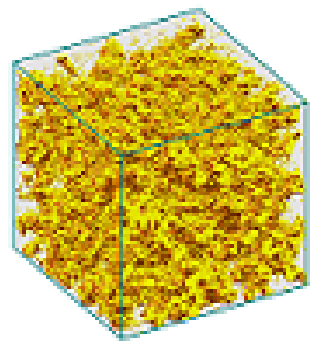} &
\includegraphics[height=5cm,keepaspectratio]{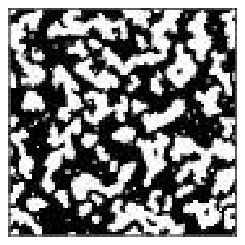} \\
\mbox{\bf (a)} & \mbox{\bf (b)}
\end{array}$
\end{center}
\caption{(color online). (a) The reconstructed 3D realization of the boron
carbide/aluminum composite from $S^{Al}_2(r)$. The aluminum phase is shown in yellow
and the ceramic phase is transparent for easy visualization. (b) A 2D slice of the
constructed 3D realization. The ceramic phase is shown in black and the aluminum
phase is shown in white. The linear size of the system $N = 160$ (pixels).}
\label{fig9}
\end{figure}

\begin{figure}[bthp]
\begin{center}
$\begin{array}{c}\\
\includegraphics[height=5.5cm,keepaspectratio]{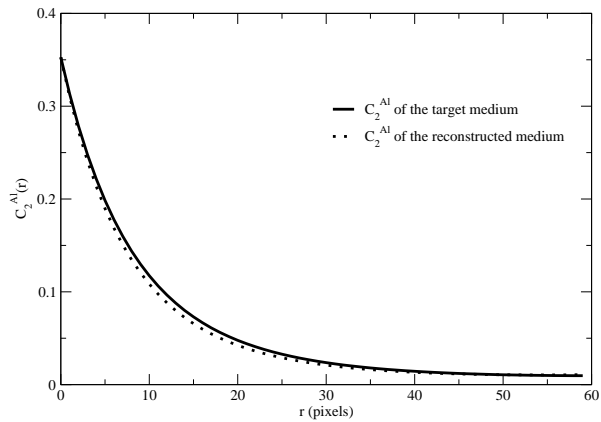}\\
\end{array}$
\end{center}
\caption{The two-point cluster functions $C^{Al}_2(r)$ for the aluminum (white) phase
 of the target and reconstructed 2D slices of the boron carbide/aluminum composite.}
\label{fig91}
\end{figure}

A 2D digitized image of a boron carbide/aluminum ($B_4C$/$Al$) inter-penetrating
composite and the sampled two-point correlation function of the \textit{aluminum}
phase (white phase) $S_2^{Al}(r)$ \cite{ceramic} are
shown in Fig.~\ref{fig8}. As we can see from Fig.~\ref{fig8}(b), $S_2^{Al}(r)$ is
essentially an exponentially decreasing function without any significant short-range
correlation. Thus, $S_2^{Al}(r)$ can be approximated by

\begin{equation}
\label{eq9}
S_2^{Al}(r) = \alpha'_1 \phi_1\phi_2 f_D(r) + \alpha'_2 \phi_1\phi_2 f_O(r) +\phi_2^2 + \delta S'_2(r),
\end{equation}

\noindent where $\alpha'_1 = 0.81$, $\alpha'_2 = 0.19$ are
combination coefficients; $\phi_1 = 0.647$ and $\phi_2 = 0.353$
are volume fractions of the boron carbide (black) phase and the
aluminum (white) phase, respectively. $f_D(r)$ and $f_O(r)$ are
given by Eq.~(\ref{eq101}) and Eq.~(\ref{eq102}), respectively;
the parameters used are $a = 3~(pixel)$, $b = 10~(pixel)$ and $q =
0.22~(pixel^{-1})$, $\psi = 0$. The average of the absolute values
of discrepancies $\Delta S'_2$ is given by

\begin{equation}
\label{eq991}
\Delta S'_2 = \frac{1}{N_L}\sum_r \left |\delta S'_2(r)\right| \simeq 6.4\times10^{-4}.
\end{equation}

\noindent where $N_L = 80$ is the sample-length.

 The 3D reconstruction of the boron carbide/aluminum composite from $S_2^{Al}(r)$
obtained from the digitized image of a 2D slice [Fig.~\ref{fig8}(a)] is shown in
Fig.~\ref{fig9}. As in the previous section, to verify that the reconstruction is
quantitatively successful, we compute and compare the two-point cluster function
$C^{Al}_2(r)$ for the aluminum phase of the target and reconstructed 2D slices of the
 composite. As shown in Fig.~\ref{fig91}, the reconstruction also slightly
underestimates $C^{Al}_2(r)$ with acceptable discrepancies (e.g., the largest
discrepancy $|\delta C^{Al}_2|\sim 8\times 10^{-3}$). The medium can be considered
as successfully reconstructed. Also note that $C^{Al}_2(r)$ is long-ranged,
indicating large clusters of the aluminum phase present in the media.

\subsubsection{Binary Laser-Speckle Pattern}

\begin{figure}[bthp]
\begin{center}
$\begin{array}{c}\\
\includegraphics[width=5cm,keepaspectratio]{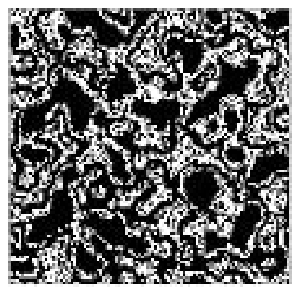}\\
\mbox{\bf (a)} \\\\\\
\includegraphics[height=5.5cm,keepaspectratio]{fig12b.eps}\\
\mbox{\bf (b)}
\end{array}$
\end{center}
\caption{(a) A digitized image of a binary laser-speckle pattern \cite{laser}. The
volume fraction of the black phase is $\phi_1 = 0.639$, and the volume fraction of
the white phase is $\phi_2 = 0.361$. (b) The two-point correlation $S^{SP}_2(r)$ of
the black phase.}
\label{fig10}
\end{figure}

Although in the above two examples our theoretical formalism works
well, there are situations where the microstructural information
contained in $S_2(r)$ alone is not sufficient to accurately
reconstruct a heterogeneous material. One such example is the
multi-scale structure of a binary laser-speckle pattern
(Fig.~\ref{fig10}). The figure reveals that there are three
structural elements: ``particles'', ``stripes'' and a background
``noise'' (individual black pixels dispersed throughout the white
phase). Thus, there are three characteristic length scales in the
medium associated with these structural elements.

\begin{figure}[bthp]
\begin{center}
$\begin{array}{c}\\
\includegraphics[width=5cm,keepaspectratio]{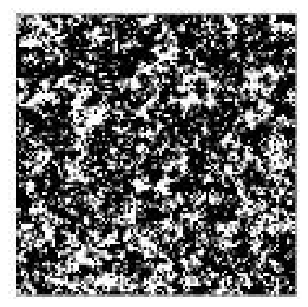}\\
\end{array}$
\end{center}
\caption{The reconstruction of the binary laser-speckle pattern
from $S^{SP}_2(r)$. The linear size of the system $N = 150$
(pixels).} \label{fig11}
\end{figure}

The reconstruction of the speckle pattern is shown in Fig.~\ref{fig11}. Comparing
Fig.~\ref{fig11} with Fig.~\ref{fig10}, we can see that instead of reproducing all
the structural elements in the target medium, the (re)construction program seems to
mix them up to generate a single-scale structure that has the same (or to a very high
 accuracy) two-point correlation function as the target medium. We note that this is a numerical
example of structural degeneracy of $S_2(r)$.

The laser-speckle pattern we considered is also an example of the
\textit{multi-scale media} (MSM). The separation of length scales in MSM
results in the inefficiency of the bulk-based structure
characteristics (e.g., $n$-point correlation
 functions), since usually they can only pick up structural information associated
with the largest length scale. For example, consider the
dislocations in a crystalline solid as the phase of interest; it
is clear that the two-point correlation function of the
``dislocation'' phase is identically zero since the
``dislocation'' phase has no measure compared with the bulk
``crystal'' phase (i.e., the volume of dislocations is zero
compared with that of the bulk crystal). In this extreme example,
the ratio $\gamma$ of the two characteristic length scales
associated with the ``dislocation'' phase $L_d$ and the
``crystal'' phase $L_c$ is zero, i.e., $\gamma = L_d/L_c = 0$. In
digitized media, the ratio of length scales is always positive due
to the discrete nature of the system; however when the ratio is
significantly different from unity, the two-point correlation
function alone is not able to capture the key structural features.
On the other hand, extensive experience with successful
reconstructions of single-scale media (SSM) from $S_2$ shows that the
two-point correlation functions are indeed sufficient to determine
the structures of SSM to a high accuracy. Thus, we conclude that
in general reconstructions using $S_2$ work well for heterogeneous
materials with single-scale structures, while the microstructural
information contained in $S_2$ is not sufficient to accurately
model multi-scale media. Note that more morphological information
(e.g., the lineal-path function \cite{linealpath} and the
two-point cluster function, etc.) can be used to model MSM more
accurately. However, even the cluster-type functions containing
connectedness information of the media such as $C_2$ can not
completely characterize MSM statistically.

\subsection{Constructing Materials with Desired Structural Characteristics}

From Eq.~(\ref{eq4}), one can construct candidates of realizable
two-point correlation functions using the basis functions. Given a
set of basis functions with diverse and interesting properties,
one would be able to construct a two-point correlation function
that exhibits all the useful properties of the basis functions to
some extend and generate an ``optimal'' structure that realizes
all the desired structural features. Thus, the theoretical
formalism enables one to design materials with \textit{structural}
properties of interest. Given an accurate
structure-property relation, one could even design materials with
\textit{physical} properties of interest by manipulating their
two-point correlation functions. For example, Adams, Gao and
Kalidindi recently developed a methodology to obtain finite
approximations of the second-order properties closure in single
phase polycrystalline materials, from which the second-order
microstructure design can proceed \cite{Adams}.

As pointed out in paper I, we know very little about the basis function set
$\{f_i(r)\}_{i=1}^m$ at this stage. Our choice of $\{f_i(r)\}_{i=1}^m$ is based on
the criteria that $f_i(r)$'s should have simple analytical forms and they present
typical features of certain known two-point correlation functions. Important features
 exhibited by most realizable two-point correlation functions are monotonically
deceasing or damped oscillating, which corresponds to materials
without or with significant short-range order, respectively.
Another feature of two-point correlation
 functions that may affect the structures of the corresponding materials is the
smoothness of the function, which is not emphasized in our previous work. In the
following, we will see through several examples that the properties of basis
functions can be observed in the generated structures; also the media with desired
structural properties can be obtained by manipulating the combination coefficients
and the parameters (e.g., effective correlation length) in the basis functions.

In the paper I, we already investigated hypothetical correlation
functions combining the exponentially decreasing and
damped-oscillating features. In this section, we provide an
example of non-smooth correlation functions (with discontinuous
derivatives), i.e., the polynomial function of order two
$f_P^{(2)}(r)$, which is defined as

\begin{equation}
\label{eq14}
f_P^{(2)}(r) = \left\{\begin{array}{*{20}c} (1-r/c)^2 \quad 0 \le r \le c, \\\\ 0 \quad\quad r>c, \end{array}\right.
\end{equation}

\noindent where the parameter $c$ is the effective correlation
length. Other correlation functions having this non-smooth feature
include the functions of known constructions (e.g., $f_S(r)$,
$f_C(r)$, etc.) and other polynomial functions
 \cite{yjiao}. A typical medium associated with this type of functions is composed of
 dispersions of fully penetrable ``particles''. The monotonically decaying part of
the functions determines the size and shape of the ``particles'' and the
penetrability is consistent with the flat ``tail'' of the functions, which implies no
 spatial correlation between the ``particles''. When these functions are use as
dominant basis functions, one can expect the generated media also contain well-defined
overlapping ``particles''; thus, surface optimization could be applied in the
 (re)constructions. However, when functions like Debye random medium function are
dominant in the convex combination, this algorithm modification should be used with
care because the media could contain ``clusters'' of all sizes and shapes, which
would significantly reduce the efficiency of the modified algorithm using surface
optimization.

Besides $f^{(2)}_P(r)$, we also use Debye random medium function $f_D(r)$ and
damped-oscillating function $f_O(r)$ as the basis functions. Consider the simplest
form of $\wp$ for these three basis functions, i.e.,

\begin{equation}
\label{eq1501}
f(r) = \alpha_1 f_D(r) + \alpha_2 f_O(r) + \alpha_3 f^{(2)}_P(r),
\end{equation}

\noindent where $\alpha_i$ ($i=1,2,3$) satisfies
$0\le\alpha_i\le1$ and $\sum\nolimits_i\alpha_i = 1$; and
$f_D(r)$, $f_O(r)$ and $f^{(2)}_P(r)$ are given by
Eq.~(\ref{eq101}), Eq.~(\ref{eq102}) and Eq.~(\ref{eq14}),
respectively. The characteristic length scales of the generated
structures are determined by the parameters in the basis
functions; and the ratios of the characteristic lengths are chosen
to be close to unity, i.e., the hypothetical media belong to SSM.

\begin{figure}[bthp]
\begin{center}
$\begin{array}{c}\\
\includegraphics[height=5.5cm,keepaspectratio]{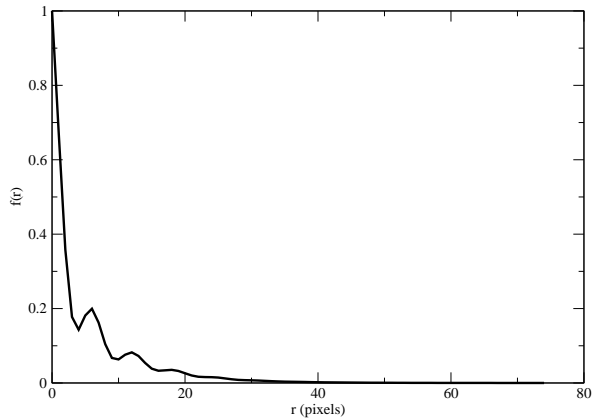}\\
\end{array}$
\end{center}
\caption{The constructed autocovariance function $f(r)$ given by Eq.~(\ref{eq1501}).
The combination coefficients are $(\alpha_1, \alpha_2, \alpha_3) = (0.3, 0.2, 0.5)$;
the parameters in the basis functions are $a=8~(pixel)$; $b=5~(pixel)$, $q=1~(pixel)^
{-1}$, $\psi = 0$; $c=5~(pixel)$.}
\label{fig12}
\end{figure}

\begin{figure}[bthp]
\begin{center}
$\begin{array}{c@{\hspace{2cm}}c}\\
\includegraphics[height=6cm,keepaspectratio]{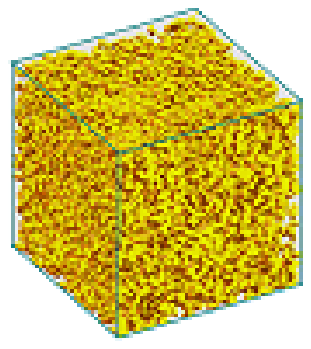} &
\includegraphics[height=5cm,keepaspectratio]{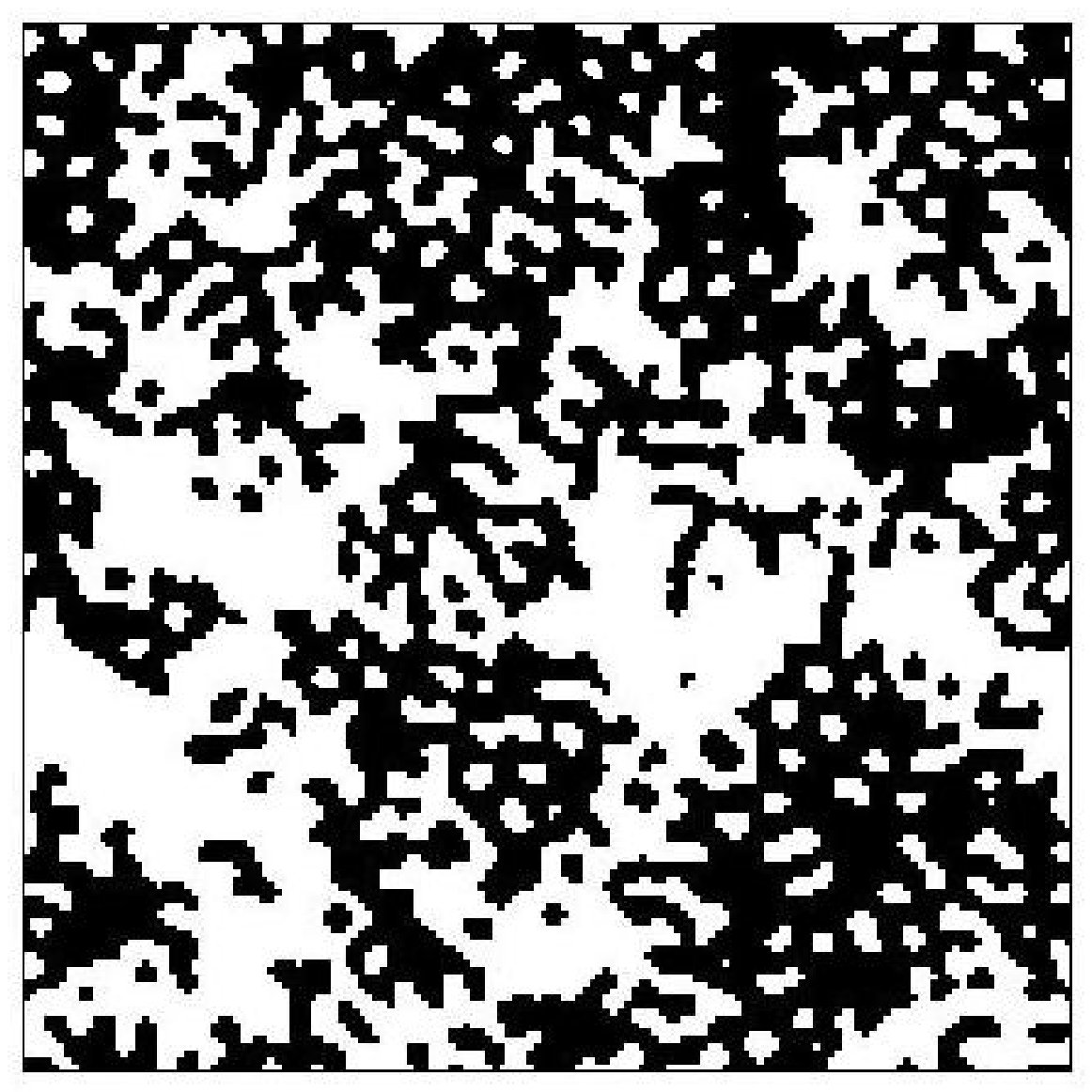} \\
\mbox{\bf (a)} & \mbox{\bf (b)}
\end{array}$
\end{center}
\caption{(color online). (a) The constructed 3D structure associated with $f(r)$
shown in Fig.~\ref{fig12} with volume fractions $\phi_1 = \phi_2 = 0.5$. The ``black
phase'' is shown in yellow and the ``white phase'' is transparent for easy
visualization. (b) A 2D slice of the constructed 3D realization. The linear size of
the system $N=150$ (pixels).}
\label{fig120}
\end{figure}

Suppose we choose the combination coefficients $(\alpha_1,
\alpha_2, \alpha_3) = (0.3, 0.2, 0.5)$; and choose the following
values for the parameters in the basis functions: $a=8~(pixel)$;
$b=5~(pixel)$, $q=1~(pixel)^{-1}$, $\psi = 0$; $c=5~(pixel)$. The
combined autocovariance function $f(r)$ is shown in
Fig.~\ref{fig12} and the constructed structure with volume
fraction of black phase $\phi_1 = 0.5$ is shown in
Fig.~\ref{fig120}. From the figures, we can see the effects of
each basis function on the generated structure. The short-range
correlations in the structure are determined by $f^{(2)}_P(r)$,
which allows spatially uncorrelated ``particles'' with diameter
$c$ to form. These ``particles'' would form clusters of different
sizes as the volume fraction of the black phase increases. The
middle-range correlations in the structure are dominated by the
oscillation part in $f(r)$, which is the contribution of $f_O(r)$.
As we have shown in paper I, the parameter $q$ is manifested as a
characteristic repulsion among different structure elements
(appearing in different forms at different volume fraction) with
diameter of order $q$; while $b$ controls the overall exponential
damping, and thus, the effective range of the repulsion. The
large-scale correlation is dominated by the long ``tail'' of
$f_D(r)$, which allows clusters of all shapes and sizes to form.
In Fig.~\ref{fig120}, we can clearly identify the structure
elements associated with each basis function. For example, the
stripe-like structures
 are associated with the oscillating feature of $f(r)$, the width of which is
approximately $q$. Several ``particles'' with diameter $a$ can be found dispersed in
the white phase, which correspond to the contribution of $f^{(2)}_P(r)$. Note that
most ``particles'' form clusters together with the ``stripes'' at this relatively
high volume fraction $\phi_1 = 0.5$. Finally, the spatial distribution of large-scale
 clusters is determined by the ``tail'' of $f_D(r)$.

\begin{figure}[bthp]
\begin{center}
$\begin{array}{c}\\
\includegraphics[height=5.5cm,keepaspectratio]{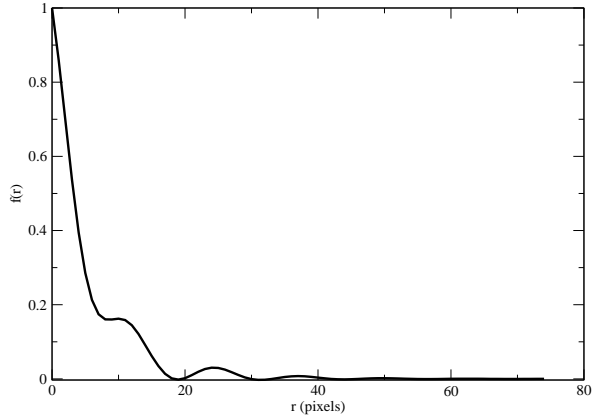}\\
\end{array}$
\end{center}
\caption{The constructed autocovariance function $f(r)$ given by
Eq.~(\ref{eq1501}). The combination coefficients are $(\alpha_1, \alpha_2, \alpha_3)
= (0.3, 0.2, 0.5)$; the parameters in the basis functions are $a=8~(pixel)$;
$b=10~(pixel)$, $q=0.5~(pixel)^{-1}$, $\psi = 0$; $c=15~(pixel)$.}
\label{fig13}
\end{figure}

\begin{figure}[bthp]
\begin{center}
$\begin{array}{c@{\hspace{2cm}}c}\\
\includegraphics[height=6cm,keepaspectratio]{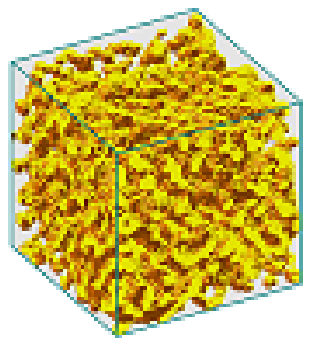} &
\includegraphics[height=5cm,keepaspectratio]{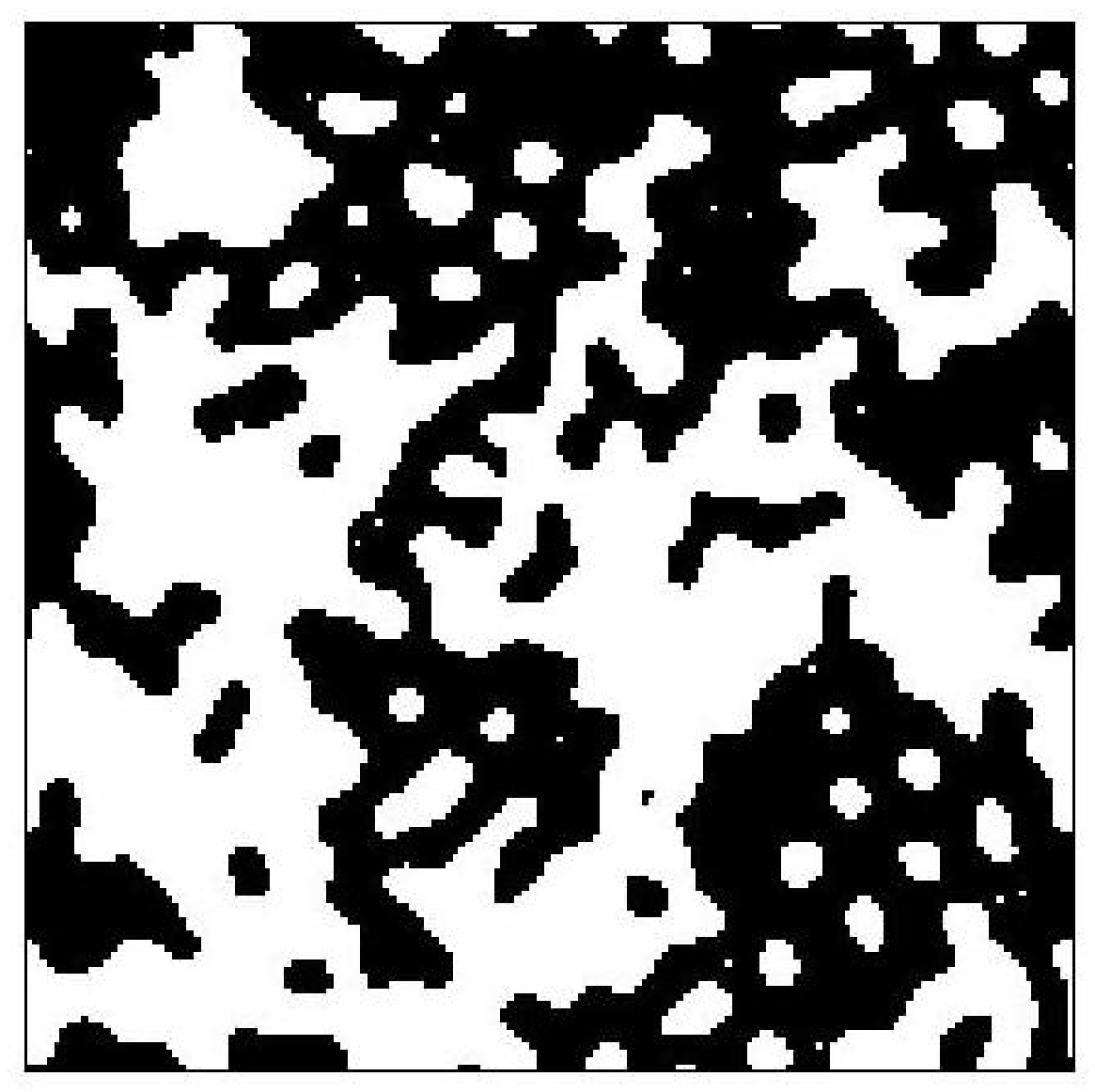} \\
\mbox{\bf (a)} & \mbox{\bf (b)}
\end{array}$
\end{center}
\caption{(color online). (a) The constructed 3D structure associated with $f(r)$
shown in Fig.~\ref{fig13} with volume fractions $\phi_1 = \phi_2 = 0.5$. The ``black
phase'' is shown in yellow and the ``white phase'' is transparent for easy
visualization. (b) A 2D slice of the constructed 3D realization. The linear size of
the system $N=150$ (pixels).}
\label{fig130}
\end{figure}

Suppose now we would like to generate a similar structure as the
one in Fig.~\ref{fig120} but with larger ``particles'' and
clusters. To ``grow'' the ``particles'', we need to increase the
parameter $c$ in $f^{(2)}_P(r)$, which controls the effective
diameter of the ``particles''. To form larger clusters, we need to
reduce the ``repulsion'' between the structural elements; or
equivalently, to suppress the oscillation introduced by $f_O(r)$.
This can be done by either increasing the parameter $b$ in
$f_O(r)$ or decreasing the coefficient $\alpha_2$. We first adopt
the former method. Thus, to construct a new structure with the
required properties, we use the following parameters:
$a=8~(pixel)$; $b=10~(pixel)$, $q=0.5~(pixel)^{-1}$, $\psi = 0$;
$c=15~(pixel)$; and use the same $\alpha_i$ $(i=1,2,3)$ as the
last example. The combined autocovariance function $f(r)$ is shown
in Fig.~\ref{fig13} and the constructed structure with volume
fraction of black phase $\phi_1 = 0.5$ is shown in
Fig.~\ref{fig130}. From the figures, we can see the generated
medium indeed contains larger ``particles'' and clusters; besides,
there are still stripe-like structures associated with the
repulsion part of $f(r)$.

\begin{figure}[bthp]
\begin{center}
$\begin{array}{c}\\
\includegraphics[height=5.5cm,keepaspectratio]{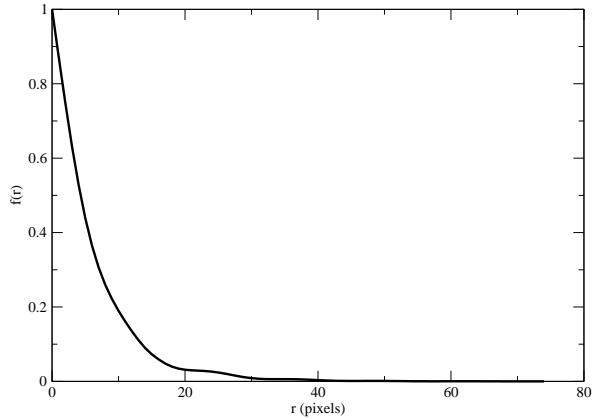}\\
\end{array}$
\end{center}
\caption{The constructed autocovariance function $f(r)$ given by
Eq.~(\ref{eq1501}). The combination coefficients are $(\alpha_1, \alpha_2, \alpha_3)
= (0.45, 0.05, 0.5)$; the parameters in the basis functions are $a=8~(pixel)$;
$b=10~(pixel)$, $q=0.5~(pixel)^{-1}$, $\psi = 0$; $c=15~(pixel)$.}
\label{fig14}
\end{figure}

\begin{figure}[bthp]
\begin{center}
$\begin{array}{c@{\hspace{2cm}}c}\\
\includegraphics[height=6cm,keepaspectratio]{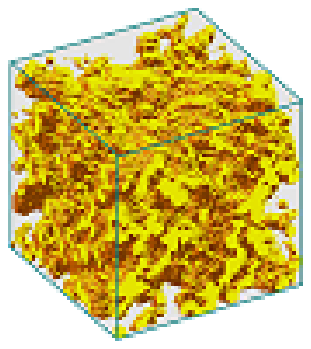} &
\includegraphics[height=5cm,keepaspectratio]{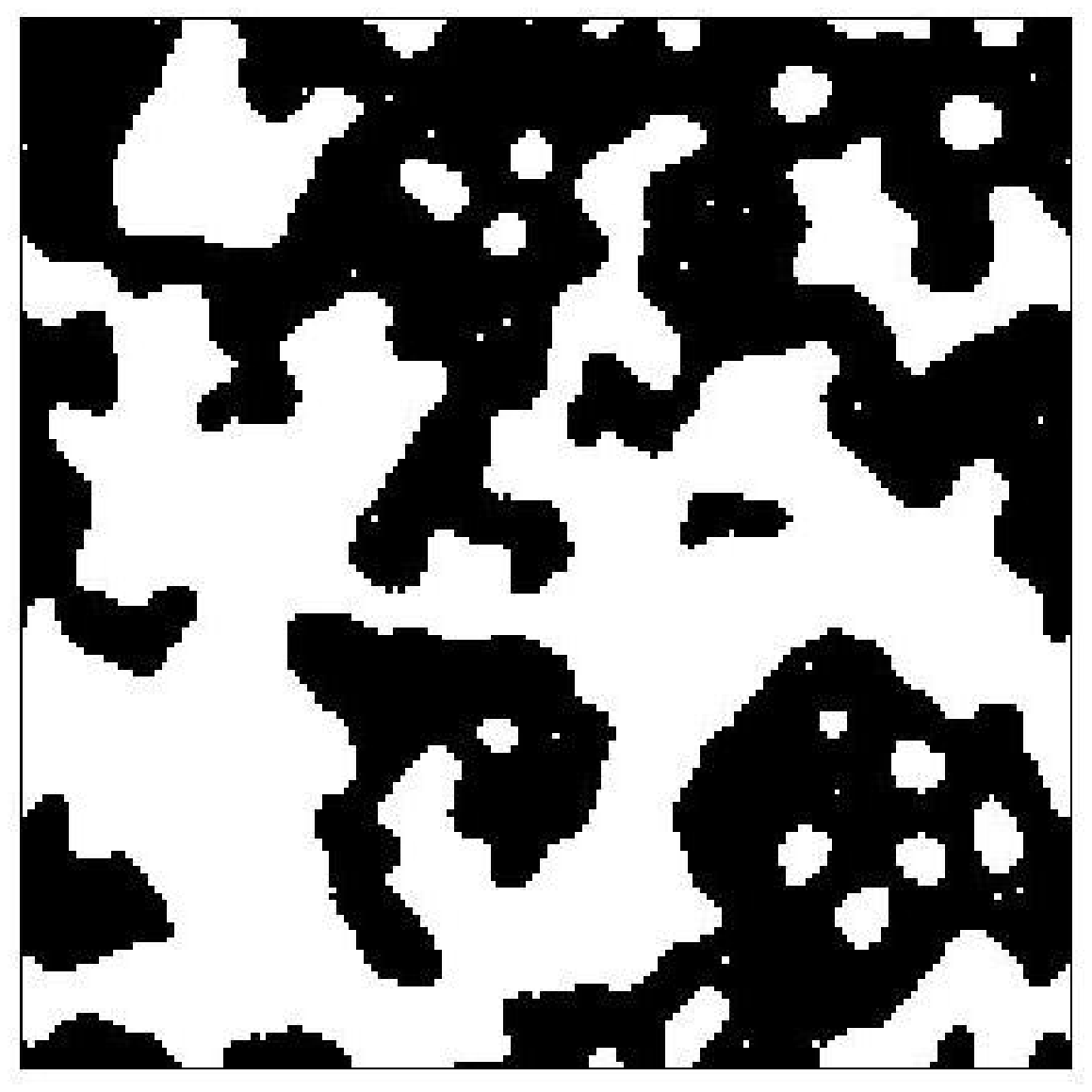} \\
\mbox{\bf (a)} & \mbox{\bf (b)}
\end{array}$
\end{center}
\caption{(color online). (a) The constructed 3D structure associated with $f(r)$
shown in Fig.~\ref{fig14} with volume fractions $\phi_1 = \phi_2 = 0.5$. The ``black
phase'' is shown in yellow and the ``white phase'' is transparent for easy
visualization. (b) A 2D slice of the constructed 3D realization. The linear size of
the system $N=150$ (pixels).}
\label{fig140}
\end{figure}

Now we would like to further increase the size of clusters in the
constructed medium while keeping all the parameters in the basis
functions unchanged. This time we need to adjust the combination
coefficients to obtain the required structure.
 To generate larger clusters, the repulsion effect need to be further suppressed;
thus, we need to reduce $\alpha_2$. Since the sum of the coefficients must equal 1,
we also need to increase the other two $\alpha_i$ ($i=1,3$). Note that further
increasing $\alpha_3$ makes $f^{(2)}_P(r)$ dominant in the combination; however, as
we discussed before, $f^{(2)}_P(r)$ corresponds to spatial uncorrelated ``particles''
 and the size of clusters formed by these ``particles'' is determined by the volume
fraction the ``particle'' phase, which cannot be changed here. So
we choose to increase $\alpha_1$ of $f_D(r)$ based on the fact
that the large correlation length and long ``tail'' of $f_D(r)$
significantly stimulate the forming of large clusters. Thus, in
the construction we use the following combination coefficients:
$(\alpha_1, \alpha_2, \alpha_3) = (0.45, 0.05, 0.5)$; and use the
same parameters in the basis functions as the last example. The
combined autocovariance function $f(r)$ is shown in
Fig.~\ref{fig14} and the constructed structure with volume
fraction of black phase $\phi_1 = 0.5$ is shown in
Fig.~\ref{fig140}.

From the above examples, we see that even a simple convex combination of the basis
functions (the simplest form of map $\wp$) can provide variety of candidates of
realizable two-point correlation functions. We can also obtain desired structures by
manipulating the combination coefficients and the parameters in the basis functions.
Thus, we argue that in general given a complete set of basis functions,
Eq.~(\ref{eq7}) enables one to construct candidates of realizable functions with
desired properties and to design materials with structural properties of interest.

\section{Conclusions}

In this paper, we described in great detail the lattice-point
algorithm, which has been shown to be both efficient and
isotropy-preserving for (re)constructing statistically homogeneous
and isotropic media. The digitized medium (pixel system) can be
considered as a lattice-gas system, in which the black pixels
behave like nonoverlapping ``particles'' moving from one lattice
site to another. The two-point correlation function of the medium
is sampled by binning all distances between black pixels and
dividing the number in each bin by the total number of distances
between all lattice sites in that bin. The energy threshold
indicating to what accuracy the medium is (re)constructed has been
discussed in detail for the first time. This quantity is directly
related to the issue of non-uniqueness of (re)constructions. We
have also described an algorithm modification using surface
optimization to further speed up the (re)construction process and
discussed its implementation based on our lattice-gas
 version of the digitized media. Numerical experiments have shown that by applying
the surface optimization algorithm properly (i.e., in the
(re)constructions of media composed well-defined ``particles''),
the final energy can be reduced by a factor of $10^{-2}$ if the
same cooling schedule is used. The choice of different
pixel-lattices has also been discussed.

We have applied the theoretical formalism proposed in paper I to
model several examples of real materials. In particular, we used
the lattice-point algorithm to generate 3D realizations of a
Fontainebleau sandstone and a boron carbide/aluminum composite
from 2D images of their slices through the materials. The two-point
cluster functions in the reconstructions in these
two different cases matched the corresponding ones
in the original materials, thus  demonstrating the efficiency
and accuracy of the (re)construction algorithm in
these cases. We also studied the reconstruction of a binary
laser-speckle pattern in 2D, in which the algorithm fails to
reproduce the target pattern accurately. We conclude that in
general reconstructions using $S_2$ only work well for
heterogeneous materials with single-scale structures. However,
two-point information via $S_2$ is not sufficient to accurately
model multi-scale media. In paper I, we pointed out that given a
complete set of basis functions, one can obtain the basis-function
approximations of the two-point correlation functions for the
materials of interest to any accuracy. Here we only used a simple
form of Eq.~(\ref{eq7}) due to our limited knowledge of the basis
functions set. We also constructed realizations of materials with
desired structural characteristics by manipulating the combination
coefficients and the parameters in the basis functions. Thus,
Eq.~(\ref{eq7}) enables one to construct candidates of realizable
functions with desired properties and to design materials with
structural properties of interest.

We are now developing efficient (re)construction procedures that
take into account additional microstructural information
that characterize the media.
For example, using our lattice-gas version of the digitized media,
the two-point cluster functions $C_2$ can
 be directly incorporated into the Yeong-Torquato scheme
\cite{2PhysRevE.57.495}, which enables one to obtain more accurate
(re)constructions of the target media. Also, a quantitative structure-property
relation is necessary for the further applications of the
theoretical formalism. Finite element analysis on the generated
media will be performed to establish the relation. Such work will
be reported in our future publications.

\begin{acknowledgments}

Acknowledgment is made to the Donors of the American
Chemical Society Petroleum Research Fund for support of
this research.

\end{acknowledgments}


\end{document}